\documentclass{article}

\usepackage[utf8]{inputenc} % allow utf-8 input
\usepackage[T1]{fontenc}    % use 8-bit T1 fonts
\usepackage{hyperref}       % hyperlinks
\usepackage{url}            % simple URL typesetting
\usepackage{booktabs}       % professional-quality tables
\usepackage{amsfonts}       % blackboard math symbols
\usepackage{nicefrac}       % compact symbols for 1/2, etc.
\usepackage{microtype}      % microtypography
\usepackage{xcolor}         % colors
\usepackage{graphicx}
\usepackage{caption}
\usepackage[final,nonatbib]{neurips_2024_ml4ps}
\usepackage{floatrow}
\usepackage[inline]{enumitem}  
\usepackage{multicol}

\title{Reconstruction of Continuous Cosmological Fields from Discrete Tracers with Graph Neural Networks}

\author{
  Yurii Kvasiuk \\
  Department of Physics, University of Wisconsin-Madison\\
  Madison, WI 53706, USA \\
  \texttt{kvasiuk@wisc.edu} \\
  \And
  Jordan Krywonos \\
  Perimeter Institute for Theoretical Physics \\
  Waterloo, ON N2L 2Y5, CA\\
  Department of Physics and Astronomy,\\
  York University\\
  Toronto, ON M3J 1P3, CA\\
  \texttt{jkrywonos@perimeterinstitute.ca} \\
  \And
  Matthew C. Johnson \\
  Perimeter Institute for Theoretical Physics \\
  Waterloo, ON N2L 2Y5, CA\\
  Department of Physics and Astronomy,\\
  York University\\
  Toronto, ON M3J 1P3, CA\\
  \texttt{mjohnson@perimeterinstitute.ca} \\
  \AND
  Moritz M\"unchmeyer \\
  Department of Physics, University of Wisconsin-Madison \\
  Madison, WI 53706, USA \\
  NSF-Simons AI Institute for the Sky (SkAI)\\
  172 E. Chestnut St., Chicago, IL 60611, USA \\
  \texttt{muenchmeyer@wisc.edu} \\
}

\begin{document}

\maketitle

\begin{abstract}
We develop a hybrid GNN-CNN architecture for the reconstruction of 3-dimensional continuous cosmological matter fields from discrete point clouds, provided by observed galaxy catalogs. Using the CAMELS hydrodynamical cosmological simulations we demonstrate that the proposed architecture allows for an accurate reconstruction of both the dark matter and electron density given observed galaxies and their features. Our approach includes a learned grid assignment scheme that improves over the traditional cloud-in-cell method. Our method can improve cosmological analyses in situations where non-luminous (and thus unobservable) continuous fields need to be estimated from luminous (observable) discrete point cloud tracers.
\end{abstract}

\section{Introduction}

The spatial distribution of cosmological fields, such as densities of dark matter and ionized gas (referred to below as electrons), carries important information about the history of the evolution of the Universe. 
For example, knowledge about the distribution of the electrons would increase the quality of large-scale velocity reconstruction with kinetic Sunyaev-Zeldovich effect \cite{Kvasiuk:2023nje} or would be helpful in understanding the nature of dark matter through the resonant conversion \cite{mondino2024axioninducedpatchyscreeningcosmic, mccarthy2024darkphotonlimitspatchy, P_rvu_2024}. More generally, knowledge of the dark matter map allows for cross-correlation studies with various probes from the electromagnetic wave spectrum, such as intensity mapping, as well as with gravitational wave sources. While being a rich source of information, continuous density fields are largely unobservable as we can only directly see the luminous matter - position and properties of galaxies - at a discrete sparse set of points in space. Hence, developing a method to efficiently reconstruct the former from the latter is an important task. On large scales, the number density of galaxies and the density of dark matter or electrons are linearly related. On small scales, non-linear gravitational evolution and baryonic interactions in the interstellar/intergalactic medium become relevant and are usually modelled with numerical simulations. Simulations can then be used to learn or approximate different aspects of underlying physical processes. 

Significant effort has recently been focused in the direction of both simulations and ML-based modeling in cosmology. To name a few, \cite{Ramanah:2020vyl, Ni:2021mzk, Zhang:2023lqi, Schanz:2023uzg,  rouhiainen2024superresolutionemulationlargecosmological} used various AI models, such as generative adversarial networks and denoising diffusion models, to increase the resolution of the cosmological simulations - a task commonly known as superresolution; \cite{Jamieson:2022lqc, Jamieson:2024fsp} developed a field-level emulator of cosmological large-scale structure; \cite{sharma2024fieldlevelemulatormodelingbaryonic} showed that the effects of baryonic physics can be emulated by a simple transfer function, applied to a (gravity-only) simulated dark matter field. %

Specifically for our task, recently in \cite{ono2024debiasingdiffusionprobabilisticreconstruction, park2023probabilisticreconstructiondarkmatter} it was shown that diffusion models are capable of reconstructing dark matter fields from observed galaxy fields. However, these works did not treat the discrete tracers as a point cloud and were only demonstrated in 2-dimensions. 
A possible approach to our reconstruction problem is to first assign the point objects to pixels and then treat the reconstruction with a field-to-field machine learning model such as a U-Net, as in \cite{Hong__2021}. However, the grid assignment is inherently sub-optimal at finite resolution. The observed late-time galaxy distribution is discrete and non-uniform on small scales, forming structures such as filaments, clusters and voids, collectively known as the cosmic web. Assigning density to a regular grid therefore results in very sparse regions where the density is low and the loss of information in regions where the density is higher than the grid resolution. On the other hand, the distribution of dark matter or electron densities is inherently continuous and can naturally be represented with a structured (regular) grid. 

The limitations of learning from unstructured discrete objects can be circumvented with Graph Neural Networks (GNN). They provide the natural representation of observed galaxy catalogs, which are in the form of tables with spatial position, redshift, luminosity and other properties. These qualities of GNNs were recently leveraged to build models capable of inferring cosmological parameters and dark matter halo masses directly from galaxy catalogs \cite{Makinen_2022, VD_hm, VD_siom}, emulating the late-time dark-matter halo distribution from N-body simulations \cite{cuestalazaro2023pointcloudapproachgenerative}, or learning baryonic properties of galaxies \cite{wu2023learninggalaxyenvironmentconnectiongraph}.
In this work, we develop a deterministic hybrid GNN-CNN model that is trained to output the continuous cosmological density fields directly from the galaxy catalogs\footnote{The code is publicly available at https://github.com/ykvasiuk/g2fnet}. We show that the proposed setup is successful in capturing non-linear information in cosmological fields and their correlation functions, both for dark matter and baryons.

\section{Methodology}
\begin{figure}[h!]
\centering
  \includegraphics[width=0.8\textwidth]{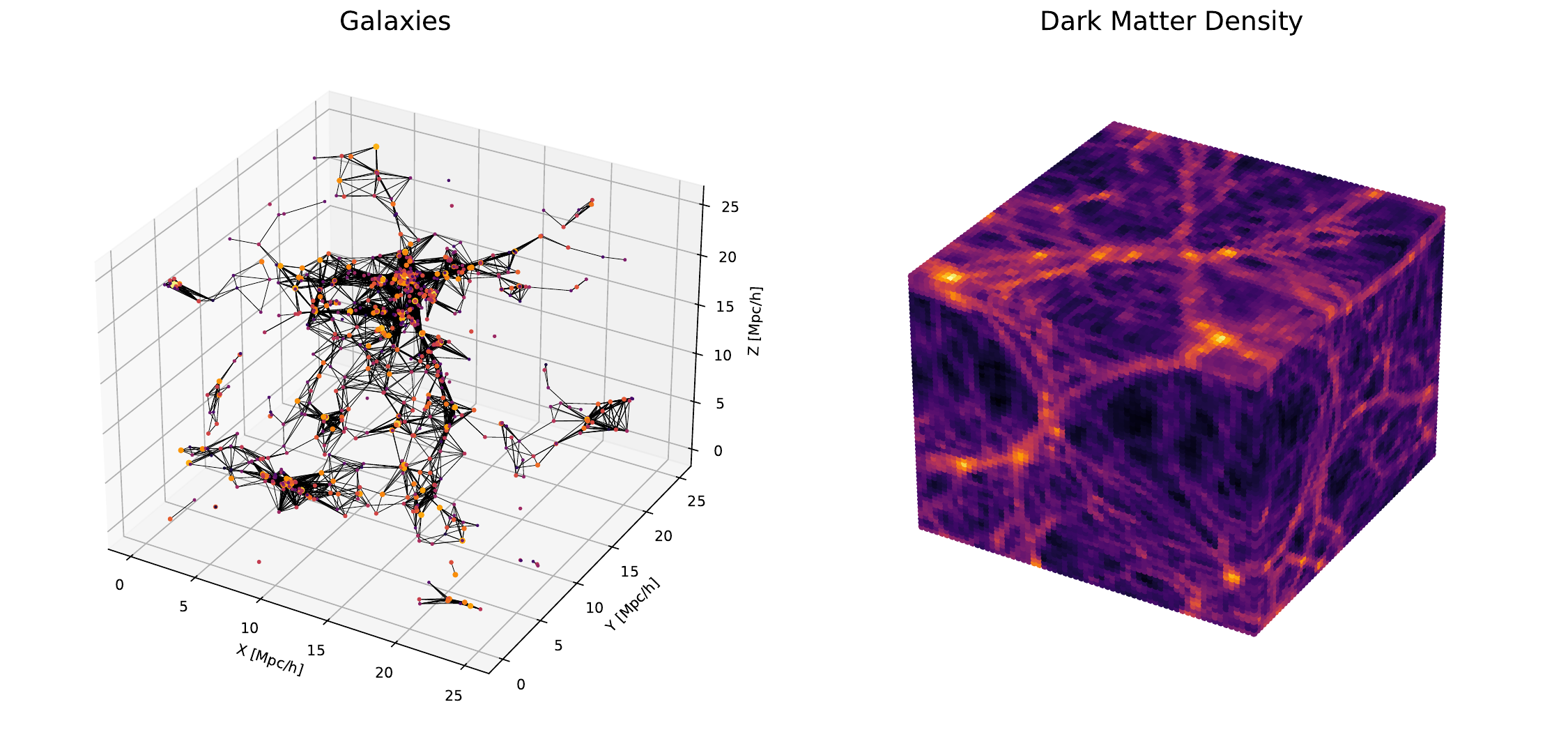}
  \caption{Visualization of the input and output. A graph constructed from the galaxy catalog (before the galaxy selection cuts) is depicted on the left ($r_{link} = 3\ \texttt{Mpc/h}$). The underlying continuous dark matter density field is shown on the right. The color scheme normalization is logarithmic. Our goal is to estimate the field on the right from the points on the left.}
    \label{fig:input_output}
\end{figure}
\subsection{Architecture}
\begin{figure}[t!]
\floatbox[{\capbeside\thisfloatsetup{capbesideposition={right,center},capbesidewidth=8cm}}]{figure}[\FBwidth]
{\caption{Grid Aggregation Procedure. Given the grid points (black) and the galaxy graph node (red dot), we choose all the grid points within a certain distance (red sphere) to the graph point and assign them the values weighted with learned radial kernel $\phi_{MLP}(r)$.}\label{fig:grid_agg}}
{\includegraphics[width=4.cm]{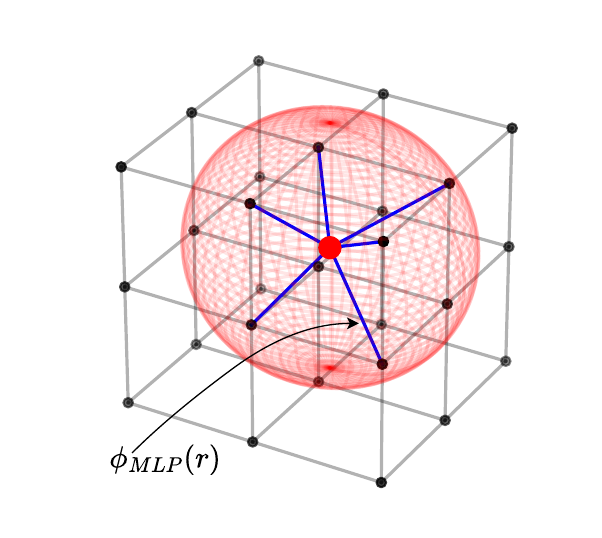}}
\end{figure}

Our proposed architecture consists of three steps and is motivated by the common encoder-decoder setup. The message-passing GNN block is used as an encoder. Then the output of the GNN block is assigned to a regular grid to form proto-density fields which then are passed to the CNN block. Since the target fields are represented by a regular spatial grid, convolutional neural nets are the natural choice for the decoder.  
For the GNN block, we use a modification of the graph message-passing scheme \cite{battaglia2018relationalinductivebiasesdeep} that also was used in \cite{VD_siom} for cosmological parameter inference from galaxy catalogs. The nodes of the graph are formed with scalar galaxy properties, the edge features are constructed from pairwise differences of positions, so that the network preserves translational symmetry. We define connectivity of the graph by providing a minimal linking length, $r_{link}$, that is one of the hyperparameters of the model. The graph might also have global features that in our case represent cosmological and astrophysical parameters. Given node features $\mathbf{w_i}$, edge features $\mathbf{e_{ij}}$, and optionally global features $u$, one message-passing step involves two operations: 
\vspace{-1em}
\begin{multicols}{2}
\begin{enumerate}
\centering
        \item $\mathbf{e^{'}_{ij}} \hookleftarrow \phi_{e}(\mathbf{w_i,w_j,\mathbf{e_{ij}}}, (u))$
        \item $\mathbf{w^{'}_i} \hookleftarrow \phi_w(\mathbf{w_i,\bigoplus_{j \in \mathcal{N}_i}\mathbf{e{'}_{ij}}}, (u))$.
\end{enumerate}
\end{multicols}
\vspace{-1em}
We parametrize $\phi_w$ and $\phi_e$ with MLPs. "$\bigoplus$" represents the permutation invariant aggregation operation that collects messages from all the edges adjacent to the node. The GNN block creates updated latent node features that we assign to the regular grid at a specified resolution. For this, we developed a grid aggregation layer that is depicted in Fig.~(\ref{fig:grid_agg}) and works as follows. The input galaxy point cloud and the grid are viewed as a bipartite graph. We assign connectivity given radius $r_{link}$ with a modified $\texttt{torch\_cluster.radius}$ function to account for the periodic boundaries. Then messages from the neighbors are aggregated to the corresponding grid points, with their contributions scaled according to a learned radial weighting kernel. This weighting kernel is derived by inputting the squared distances between the galaxy point cloud and the grid locations into a MLP with three fully connected layers and two ReLU activation functions. Then the formed fields are passed through the CNN-decoder that we represent with a UNet~\cite{ronneberger2015unetconvolutionalnetworksbiomedical}. We employ skip-connection in the convolutional layers of the UNet encoder. We also use circular-padded convolutions to account for the periodic structure of the box.

To summarize, given initial scalar galaxy properties $\mathbf{s^{0}_i}$, coordinates $\mathbf{x_i}$, and optional global parameters $u$, the target densities are formed according to the following algorithm: \\
\vspace{-1.5em}
\begin{center}
\begin{enumerate*}
    \item  \hspace{0.25em} $\mathbf{w'_i} = \texttt{mpgnn}(\mathbf{s^{0}_i},\mathbf{e^{0}_{ij}}, (u))$ \hspace{0.25em}
    \item \hspace{0.25em} $\mathbf{\delta}^{in} = \texttt{grid\_aggregate}(\mathbf{w^{'}_i},\mathbf{x_i}) $ \hspace{0.25em}
    \item \hspace{0.25em} $\mathbf{\hat{\delta}} = \texttt{unet}(\mathbf{\delta}^{in})$. \hspace{0.25em}
\end{enumerate*}
\end{center}

\subsection{Dataset}
For training, we use snapshots and SUBFIND subhalo/galaxy catalogs of IllustrisTNG-LH suite of CAMELS \cite{CAMELS_DR1,CAMELS_DR2,CAMELS_presentation} simulation set at redshift $z=0$. Each simulation follows the evolution of $256^3$ dark matter and $256^3$ gas particles in a box with a length of $25\ \texttt{Mpc/h}$. There are 1000 simulations in total from this subset. They have variable cosmological parameters $\Omega_m$ and $\sigma_8$, as well as astrophysical parameters $A_{SN1,2}$, $A_{GN1,2}$ that represent the effects of stellar winds and the influence of the active galactic nuclei. We use 27 simulations from the CV subset to evaluate the performance of the trained models. These simulations have fixed cosmological parameters and differ only by initial random seed. 

\subsection{Training}
We are constructing the simulated input galaxy catalogs from FOF-Subfind subhalo catalogs of CAMELS-IllustrisTNG simulations. We follow the selection criteria of \cite{Wu_2024}:
\begin{itemize}
    \item Both stellar and dark matter half-mass radii are larger than two Plummer radii ($R_p=0.74\ \texttt{kpc}$).
    \item The number of both stellar and dark matter particles is larger than 200.
    \item Every subhalo is assumed to be a separate galaxy.
\end{itemize}
After applying the selection cuts, we have $\mathcal{O}(200)$ galaxies in a $25^3\ (\texttt{Mpc/h})^3$ volume. While not a fully realistic galaxy data set, our selection is broadly compatible with the data from ongoing and future high-density galaxy surveys. The number density of galaxies after the cuts in our data set is approximately $10$ times higher than the one of DESI at $z=1$, but somewhat below the expected number density of Rubin Observatory \cite{M_nchmeyer_2019, Hadzhiyska_2023}. 

We then generate the graph corresponding to the galaxy catalogue. The graph nodes are generated from scalar features such as stellar mass, or velocity dispersion. We construct translationally invariant graph edge features from pairwise differences of vector quantities, like positions. The complete set of features used and their normalization are listed in Appendix \ref{app:a}. We found that it is sufficient to have one message-passing layer with $r_{link}=2 \ \texttt{Mpc/h}$. We experimented with larger values of $r_{link}$ but didn't find a significant improvement. The targets are the corresponding electron and dark matter overdensities, $\delta_e$ and $\delta_m$, that we put on a downsampled grid of $128^3$ pixels. We use the AdamW \cite{loshchilov2019decoupledweightdecayregularization} optimizer with the learning rate $\texttt{lr}=2 \times 10^{-3}$ and weight decay $\texttt{wd}=2 \times 10^{-2}$ to minimize ordinary $l_1$ loss (which we found to perform better than $l_2$):
\begin{equation}
    l_1 = \sum_{f=\delta_e,\delta_m} \left|\hat{f}-f^{true}\right|.
\end{equation}
Here, $\hat{f}$ stands for the output of the model. Out of 1000 simulations of LH suite, we use 850 first ones for training and keep 150 last ones to track the validation loss. We train two versions of the model, one that doesn't know about the true simulation parameters and the other that has access to them explicitly in the form of global features of the graph. More precisely, let us define $P(\mathbf{f},\theta|\texttt{g})$ - a probability of the true density fields $\mathbf{f}$ and parameters $\theta$ given the observed galaxy cloud $\texttt{g}$, represented as a graph. In the first case, we learn the median of the marginalized (over the parameters) probability density function
\begin{equation}
    P(\mathbf{f}|\texttt{g}) = \int d\theta P(\mathbf{f},\theta|\texttt{g}).
\end{equation}
In the second scenario, we learn the median of
\begin{equation}
    P(\mathbf{f}, \theta = \theta^* |\texttt{g}) \propto P(\mathbf{f}|\texttt{g},\theta^*).
\end{equation}
\subsection{Performance metrics}
We evaluate the cross-correlation coefficient of the Fourier modes as a function of the magnitude of a wave vector $k$ to evaluate the performance of our model. It is defined as follows:
\begin{equation}
\label{eq:crosscorr}
    r^2(k)=\frac{\langle X_{true}\hat{X} \rangle^2}{\langle X_{true}^2\rangle\langle \hat{X}^2\rangle}.
\end{equation}
The cross-correlation coefficient is the relevant quantity for most practical applications, which involve cross-correlation of different data sets. 

\section{Results}

As discussed, to isolate the effects of variable cosmological and astrophysical parameters from the ability of the model to learn, we trained two different models with and without known parameters. As a baseline to compare the cross-correlation to, we took a stellar density field $\delta^*$ that we computed by weighting galaxies by their stellar mass and gridding them with the cloud-in-cell (CIC) procedure. This quantity is directly observable through stellar luminosity and correlates well with the true dark matter distribution on large scales. If treated as a tracer of dark matter, it would assume that the amount of dark matter is proportional to the mass of the observed galaxy which is located at its position. That's not the case on smaller length scales and we expect our machine learning model to perform better there. Indeed, as we see from Fig.~(\ref{fig:rk_dm_de}) (and Fig.~(\ref{fig:snrk_dm_de}) in the Appendix \ref{app:b}), the neural network is able to reconstruct both dark matter and electron densities from the galaxy features at high fidelity. The model is noticeably better for dark-matter density, though. The reason for this is that the distribution of galaxies itself is better correlated with the underlying dark matter density. The effect of known vs. unknown cosmological and astrophysical parameters manifests itself in marginally improving the mean cross-correlation coefficient for electron density reconstruction. This surprising lack of improvement (compared to the unconditional case) can be attributed to the fact that the neural net is able to learn the cosmological and astrophysical parameters implicitly in the case where we do not provide them. Generally, our results are encouraging for practical applications. For example, for kSZ velocity reconstruction \cite{Deutsch:2017ybc,Smith:2018bpn, Cayuso:2021ljq}, the expected noise of the observable scales as $N \propto r^{-2}$ (\cite{Kvasiuk:2023nje}) where $r$ is the reconstruction coefficient of the baryon density. Comparing to Fig. \ref{fig:rk_dm_de}, this can be up to a factor of 2 improvement in the relevant $k$-range. A visualization of model input and output is shown in Fig. \ref{fig:model_otp_compare2} and \ref{fig:model_otp_compare1} of Appendix \ref{app:viz}.

\begin{figure}[t!]
\centering
  \includegraphics[width=0.9\linewidth]{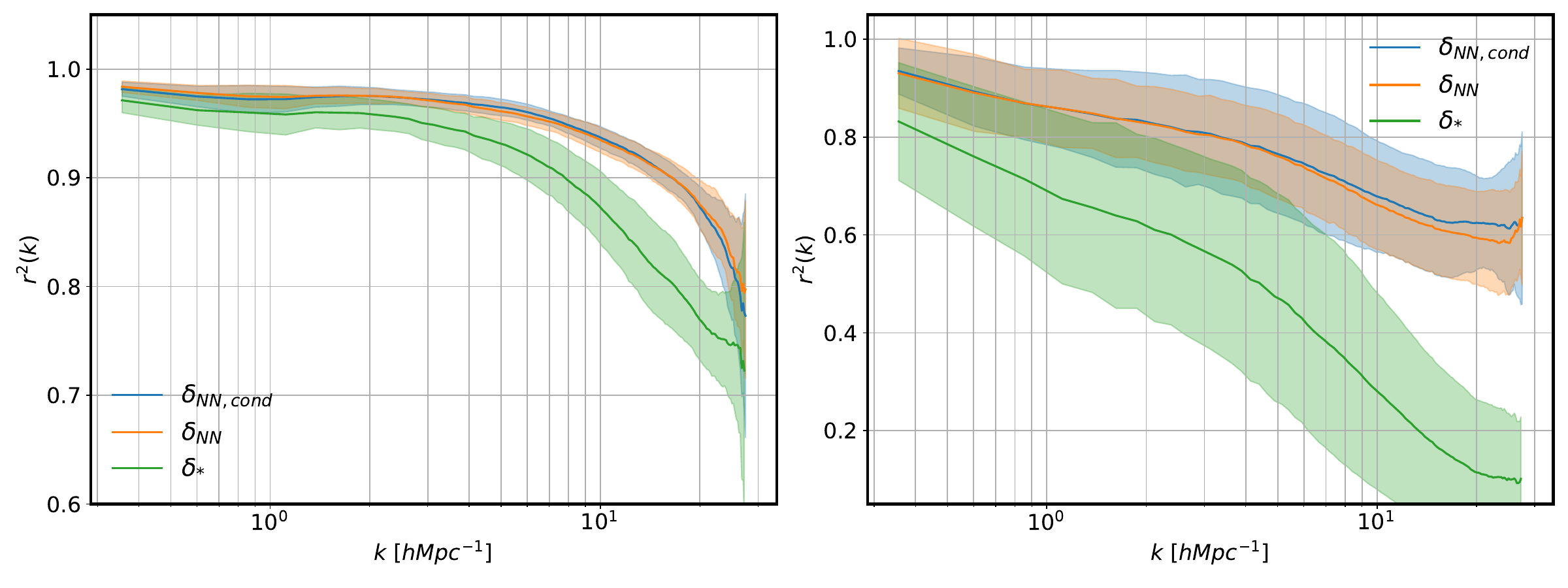}
  \caption{Left: The cross-correlation coefficient Eq. \ref{eq:crosscorr} of the reconstructed conditional and marginalized dark matter density, $\delta_{NN, cond}$ (blue) and $\delta_{NN}$ (orange) correspondingly, and stellar mass field $\delta_{*}$ (green) with a true dark matter density field as a function of wave vector $k$. Right: The same but for the electron density field. The shaded region indicates a 1-$\sigma$ contour.}
  \label{fig:rk_dm_de}
\end{figure}

\section{Conclusions and Outlook}

We've designed a hybrid GNN-CNN-based model for the reconstruction of cosmological fields directly from galaxy catalogs. The benefit of this approach is that there is no need for a pre-processing step of gridding the catalog. The graph structure allows a more straightforward way to incorporate the galaxy features directly as node attributes and positional and kinematic variables as edge features. The proposed approach nicely bridges the gap between the discreteness of observable tracers and the continuous nature of a field-level analysis and opens new possibilities for various simulation- or forward-modeling-based inference schemes directly on the catalog level.  

There are several interesting future directions. First,  it's important to consider a more realistic scenario, where we also incorporate the corresponding uncertainties of the observed quantities. One way to do so is to add some noise to the node and edge features. An additional level of realism is to add a survey mask and redshift evolution. It would be very interesting to forecast the performance of our method for Rubin Observatory. The very high galaxy number density of this experiment could benefit our method, while its photometric redshift errors may decrease the performance. Another important aspect is robustness. It would be useful to cross-check the performance on other hydrodynamical simulations with different subgrid models. Fortunately, for some applications (in particular squeezed limit observables), baryonic uncertainty in the machine learning model of the small-scale field can be handled by a bias parameter \cite{Kvasiuk:2023nje}. 
One could also perform a probabilistic rather than deterministic reconstruction. The proposed architecture can be used as a conditional encoder for a denoising diffusion or a normalizing flow architecture. In that way, one would reconstruct the whole conditional probability density of the true field given the observed one. In two dimensions at field level this was recently done in \cite{ono2024debiasingdiffusionprobabilisticreconstruction, park2023probabilisticreconstructiondarkmatter}. However, in many situations in cosmology, sampling over the conditional probability density is computationally prohibitive and a deterministic approach, as presented here, is sufficient.

\paragraph{Acknowledgments}

We thank H. Ganjoo for discussions and collaboration on alternative architectures for reconstructing cosmological fields. We thank C. K. Jespersen for discussions on graph neural networks.  M.M. acknowledges the support by the U.S. Department of Energy, Office of Science, Office of High Energy Physics under Award Number DE-SC-0017647, and by the National Science Foundation (NSF) under Grant Number 2307109. M.C.J. is supported by the National Science and Engineering Research Council through a Discovery grant. J.K. acknowledges support from the Natural Sciences and Engineering Research Council of Canada (NSERC) through the Vanier Canada Graduate Scholarship. Y.K. and M.M. are grateful for the hospitality of Perimeter Institute where a part of this work was done. Research at Perimeter Institute is supported in part by the Government of Canada through the Department of Innovation, Science and Economic Development Canada and by the Province of Ontario through the Ministry of Colleges and Universities. 

\newpage
%\printbibliography
\bibliographystyle{unsrt}
\bibliography{References}

\begin{thebibliography}{10}

\bibitem{Kvasiuk:2023nje}
Yurii Kvasiuk and Moritz M\"unchmeyer.
\newblock {Autodifferentiable likelihood pipeline for the cross-correlation of CMB and large-scale structure due to the kinetic Sunyaev-Zeldovich effect}.
\newblock {\em Phys. Rev. D}, 109(8):083515, 2024.

\bibitem{mondino2024axioninducedpatchyscreeningcosmic}
Cristina Mondino, Dalila Pîrvu, Junwu Huang, and Matthew~C. Johnson.
\newblock Axion-induced patchy screening of the cosmic microwave background, 2024.

\bibitem{mccarthy2024darkphotonlimitspatchy}
Fiona McCarthy, Dalila Pirvu, J.~Colin Hill, Junwu Huang, Matthew~C. Johnson, and Keir~K. Rogers.
\newblock Dark photon limits from patchy dark screening of the cosmic microwave background, 2024.

\bibitem{P_rvu_2024}
Dalila Pîrvu, Junwu Huang, and Matthew~C. Johnson.
\newblock Patchy screening of the cmb from dark photons.
\newblock {\em Journal of Cosmology and Astroparticle Physics}, 2024(01):019, January 2024.

\bibitem{Ramanah:2020vyl}
Doogesh~Kodi Ramanah, Tom Charnock, Francisco Villaescusa-Navarro, and Benjamin~D. Wandelt.
\newblock {Super-resolution emulator of cosmological simulations using deep physical models}.
\newblock {\em Mon. Not. Roy. Astron. Soc.}, 495:4227, 2020.

\bibitem{Ni:2021mzk}
Yueying Ni, Yin Li, Patrick Lachance, Rupert A.~C. Croft, Tiziana Di~Matteo, Simeon Bird, and Yu~Feng.
\newblock {AI-assisted superresolution cosmological simulations \textendash{} II. Halo substructures, velocities, and higher order statistics}.
\newblock {\em Mon. Not. Roy. Astron. Soc.}, 507(1):1021--1033, 2021.

\bibitem{Zhang:2023lqi}
Xiaowen Zhang, Patrick Lachance, Yueying Ni, Yin Li, Rupert A.~C. Croft, Tiziana Di~Matteo, Simeon Bird, and Yu~Feng.
\newblock {AI-assisted super-resolution cosmological simulations III: time evolution}.
\newblock {\em Mon. Not. Roy. Astron. Soc.}, 528(1):281--293, 2024.

\bibitem{Schanz:2023uzg}
Andreas Schanz, Florian List, and Oliver Hahn.
\newblock {Stochastic Super-resolution of Cosmological Simulations with Denoising Diffusion Models}.
\newblock 10 2023.

\bibitem{rouhiainen2024superresolutionemulationlargecosmological}
Adam Rouhiainen, Michael Gira, Moritz Münchmeyer, Kangwook Lee, and Gary Shiu.
\newblock Super-resolution emulation of large cosmological fields with a 3d conditional diffusion model, 2024.

\bibitem{Jamieson:2022lqc}
Drew Jamieson, Yin Li, Renan~Alves de~Oliveira, Francisco Villaescusa-Navarro, Shirley Ho, and David~N. Spergel.
\newblock {Field-level Neural Network Emulator for Cosmological N-body Simulations}.
\newblock {\em Astrophys. J.}, 952(2):145, 2023.

\bibitem{Jamieson:2024fsp}
Drew Jamieson, Yin Li, Francisco Villaescusa-Navarro, Shirley Ho, and David~N. Spergel.
\newblock {Field-level Emulation of Cosmic Structure Formation with Cosmology and Redshift Dependence}.
\newblock 8 2024.

\bibitem{sharma2024fieldlevelemulatormodelingbaryonic}
Divij Sharma, Biwei Dai, Francisco Villaescusa-Navarro, and Uros Seljak.
\newblock A field-level emulator for modeling baryonic effects across hydrodynamic simulations, 2024.

\bibitem{ono2024debiasingdiffusionprobabilisticreconstruction}
Victoria Ono, Core~Francisco Park, Nayantara Mudur, Yueying Ni, Carolina Cuesta-Lazaro, and Francisco Villaescusa-Navarro.
\newblock Debiasing with diffusion: Probabilistic reconstruction of dark matter fields from galaxies with camels, 2024.

\bibitem{park2023probabilisticreconstructiondarkmatter}
Core~Francisco Park, Victoria Ono, Nayantara Mudur, Yueying Ni, and Carolina Cuesta-Lazaro.
\newblock Probabilistic reconstruction of dark matter fields from biased tracers using diffusion models, 2023.

\bibitem{Hong__2021}
Sungwook~E. Hong, Donghui Jeong, Ho~Seong~Hwang, and Juhan Kim.
\newblock Revealing the local cosmic web from galaxies by deep learning.
\newblock {\em The Astrophysical Journal}, 913(1):76, May 2021.

\bibitem{Makinen_2022}
T~Lucas Makinen, Tom Charnock, Pablo Lemos, Natalie Porqueres, Alan~F Heavens, and Benjamin~D Wandelt.
\newblock The cosmic graph: Optimal information extraction from large-scale structure using catalogues.
\newblock {\em The Open Journal of Astrophysics}, 5(1), December 2022.

\bibitem{VD_hm}
Pablo Villanueva-Domingo, Francisco Villaescusa-Navarro, Daniel Anglés-Alcázar, Shy Genel, Federico Marinacci, David~N. Spergel, Lars Hernquist, Mark Vogelsberger, Romeel Dave, and Desika Narayanan.
\newblock Inferring halo masses with graph neural networks.
\newblock {\em The Astrophysical Journal}, 935(1).

\bibitem{VD_siom}
Pablo Villanueva-Domingo and Francisco Villaescusa-Navarro.
\newblock Learning cosmology and clustering with cosmic graphs.
\newblock {\em The Astrophysical Journal}, 937(2):115, October 2022.

\bibitem{cuestalazaro2023pointcloudapproachgenerative}
Carolina Cuesta-Lazaro and Siddharth Mishra-Sharma.
\newblock A point cloud approach to generative modeling for galaxy surveys at the field level, 2023.

\bibitem{wu2023learninggalaxyenvironmentconnectiongraph}
John~F. Wu and Christian~Kragh Jespersen.
\newblock Learning the galaxy-environment connection with graph neural networks, 2023.

\bibitem{battaglia2018relationalinductivebiasesdeep}
Peter~W. Battaglia, Jessica~B. Hamrick, Victor Bapst, Alvaro Sanchez-Gonzalez, Vinicius Zambaldi, Mateusz Malinowski, Andrea Tacchetti, David Raposo, Adam Santoro, Ryan Faulkner, Caglar Gulcehre, Francis Song, Andrew Ballard, Justin Gilmer, George Dahl, Ashish Vaswani, Kelsey Allen, Charles Nash, Victoria Langston, Chris Dyer, Nicolas Heess, Daan Wierstra, Pushmeet Kohli, Matt Botvinick, Oriol Vinyals, Yujia Li, and Razvan Pascanu.
\newblock Relational inductive biases, deep learning, and graph networks, 2018.

\bibitem{ronneberger2015unetconvolutionalnetworksbiomedical}
Olaf Ronneberger, Philipp Fischer, and Thomas Brox.
\newblock U-net: Convolutional networks for biomedical image segmentation, 2015.

\bibitem{CAMELS_DR1}
Francisco {Villaescusa-Navarro}, Shy {Genel}, Daniel {Angl{\'e}s-Alc{\'a}zar}, Lucia~A. {Perez}, Pablo {Villanueva-Domingo}, Digvijay {Wadekar}, Helen {Shao}, Faizan~G. {Mohammad}, Sultan {Hassan}, Emily {Moser}, Erwin~T. {Lau}, Luis~Fernando {Machado Poletti Valle}, Andrina {Nicola}, Leander {Thiele}, Yongseok {Jo}, Oliver H.~E. {Philcox}, Benjamin~D. {Oppenheimer}, Megan {Tillman}, ChangHoon {Hahn}, Neerav {Kaushal}, Alice {Pisani}, Matthew {Gebhardt}, Ana~Maria {Delgado}, Joyce {Caliendo}, Christina {Kreisch}, Kaze W.~K. {Wong}, William~R. {Coulton}, Michael {Eickenberg}, Gabriele {Parimbelli}, Yueying {Ni}, Ulrich~P. {Steinwandel}, Valentina {La Torre}, Romeel {Dave}, Nicholas {Battaglia}, Daisuke {Nagai}, David~N. {Spergel}, Lars {Hernquist}, Blakesley {Burkhart}, Desika {Narayanan}, Benjamin {Wandelt}, Rachel~S. {Somerville}, Greg~L. {Bryan}, Matteo {Viel}, Yin {Li}, Vid {Irsic}, Katarina {Kraljic}, Federico {Marinacci}, and Mark {Vogelsberger}.
\newblock {The CAMELS Project: Public Data Release}.
\newblock {\em Astrophys. J. Suppl.}, 265(2):54, April 2023.

\bibitem{CAMELS_DR2}
Yueying {Ni}, Shy {Genel}, Daniel {Angl{\'e}s-Alc{\'a}zar}, Francisco {Villaescusa-Navarro}, Yongseok {Jo}, Simeon {Bird}, Tiziana {Di Matteo}, Rupert {Croft}, Nianyi {Chen}, Natal{\'\i} S.~M. {de Santi}, Matthew {Gebhardt}, Helen {Shao}, Shivam {Pandey}, Lars {Hernquist}, and Romeel {Dave}.
\newblock {The CAMELS Project: Expanding the Galaxy Formation Model Space with New ASTRID and 28-parameter TNG and SIMBA Suites}.
\newblock {\em Astrophys. J.}, 959(2):136, December 2023.

\bibitem{CAMELS_presentation}
Francisco {Villaescusa-Navarro}, Daniel {Angl{\'e}s-Alc{\'a}zar}, Shy {Genel}, David~N. {Spergel}, Rachel~S. {Somerville}, Romeel {Dave}, Annalisa {Pillepich}, Lars {Hernquist}, Dylan {Nelson}, Paul {Torrey}, Desika {Narayanan}, Yin {Li}, Oliver {Philcox}, Valentina {La Torre}, Ana {Maria Delgado}, Shirley {Ho}, Sultan {Hassan}, Blakesley {Burkhart}, Digvijay {Wadekar}, Nicholas {Battaglia}, Gabriella {Contardo}, and Greg~L. {Bryan}.
\newblock {The CAMELS Project: Cosmology and Astrophysics with Machine-learning Simulations}.
\newblock {\em Astrophys. J.}, 915(1):71, July 2021.

\bibitem{Wu_2024}
Sirui Wu, Nicola~R. Napolitano, Crescenzo Tortora, Rodrigo von Marttens, Luciano Casarini, Rui Li, and Weipeng Lin.
\newblock Total and dark mass from observations of galaxy centers with machine learning.
\newblock {\em Astronomy \& Astrophysics}, 686:A80, May 2024.

\bibitem{M_nchmeyer_2019}
Moritz Münchmeyer, Mathew~S. Madhavacheril, Simone Ferraro, Matthew~C. Johnson, and Kendrick~M. Smith.
\newblock Constraining local non-gaussianities with kinetic sunyaev-zel’dovich tomography.
\newblock {\em Physical Review D}, 100(8), October 2019.

\bibitem{Hadzhiyska_2023}
Boryana Hadzhiyska, Lars Hernquist, Daniel Eisenstein, Ana~Maria Delgado, Sownak Bose, Rahul Kannan, Rüdiger Pakmor, Volker Springel, Sergio Contreras, Monica Barrera, Fulvio Ferlito, César Hernández-Aguayo, Simon D~M White, and Carlos Frenk.
\newblock The millenniumtng project: refining the one-halo model of red and blue galaxies at different redshifts.
\newblock {\em Monthly Notices of the Royal Astronomical Society}, 524(2):2524–2538, July 2023.

\bibitem{loshchilov2019decoupledweightdecayregularization}
Ilya Loshchilov and Frank Hutter.
\newblock Decoupled weight decay regularization, 2019.

\bibitem{Deutsch:2017ybc}
Anne-Sylvie Deutsch, Emanuela Dimastrogiovanni, Matthew~C. Johnson, Moritz M\"unchmeyer, and Alexandra Terrana.
\newblock {Reconstruction of the remote dipole and quadrupole fields from the kinetic Sunyaev Zel\textquoteright{}dovich and polarized Sunyaev Zel\textquoteright{}dovich effects}.
\newblock {\em Phys. Rev. D}, 98(12):123501, 2018.

\bibitem{Smith:2018bpn}
Kendrick~M. Smith, Mathew~S. Madhavacheril, Moritz M\"unchmeyer, Simone Ferraro, Utkarsh Giri, and Matthew~C. Johnson.
\newblock {KSZ tomography and the bispectrum}.
\newblock 2018.

\bibitem{Cayuso:2021ljq}
Juan Cayuso, Richard Bloch, Selim~C. Hotinli, Matthew~C. Johnson, and Fiona McCarthy.
\newblock {Velocity reconstruction with the cosmic microwave background and galaxy surveys}.
\newblock {\em JCAP}, 02:051, 2023.

\end{thebibliography}

%%%%%%%%%%%%%%%%%%%%%%%%%%%%%%%%%%%%%%%%%%%%%%%%%%%%%%%%%%%%

\appendix

\section{Full set of input features and normalization}
\label{app:a}

We used the following galaxy features for training:  
\begin{itemize}
    \item $\vec{x}$ - Galaxy position
    \item $r,g,z$ - Magnitudes of r,g, and z luminosity bands 
    \item $r_{1/2}$ - Stellar half-mass radius
    \item $M_{*}$ - Stellar mass
    \item $\sigma_v$ - Velocity dispersion 
\end{itemize}
Coordinates are normalized by the sidelength of the box and vary in the unit range. Magnitudes of the luminosity bands are normalized to have zero mean and unit variance. All other features - stellar mass, stellar half-mass radius, and velocity dispersion - are log-normalized, so that $\log_{10}$ of the corresponding quantity has zero mean and the variance of one. 

\section{Signal-to-noise ratio}
\label{app:b}
\begin{figure}[tbh!]
\centering
  \includegraphics[width=0.95\linewidth]{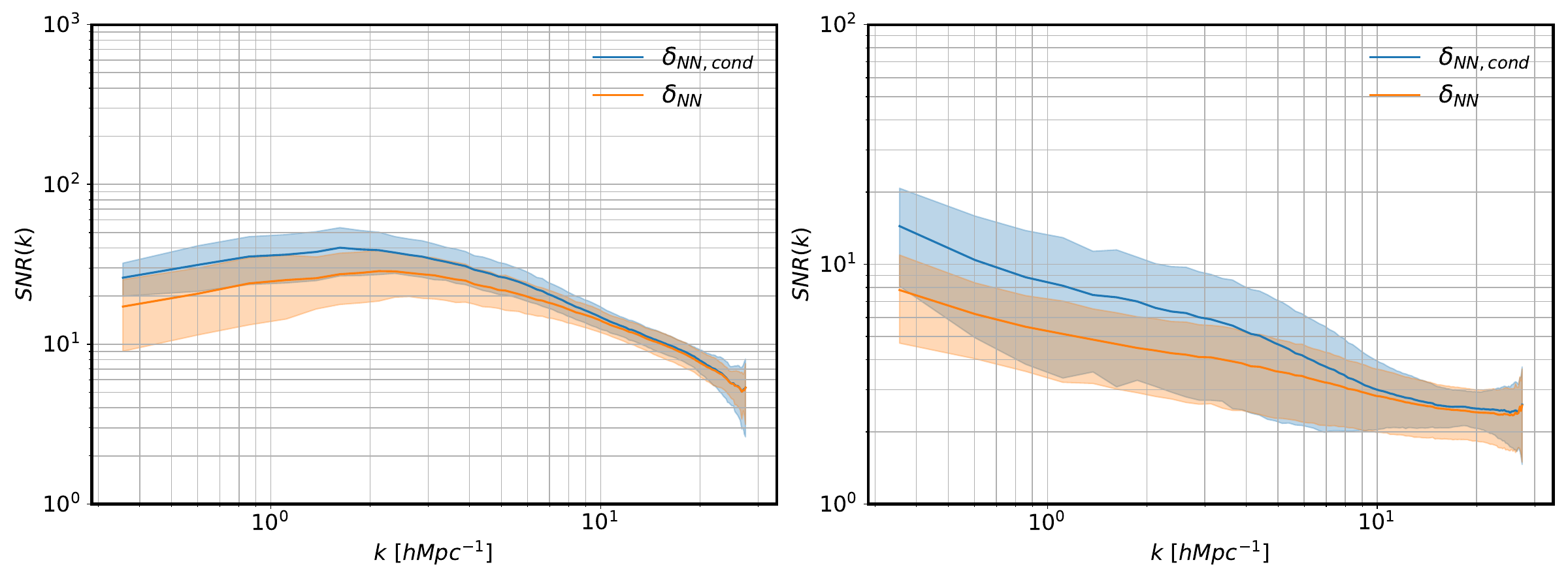}
  \caption{Left: The signal-to-noise ratio of the reconstructed conditional and marginalized dark matter density $\delta_{NN, cond}$ (blue), $\delta_{NN}$ (orange) correspondingly. Right: The same but for the electron density field.}
  \label{fig:snrk_dm_de}
\end{figure}

Another useful quality measure is the signal-to-noise ratio of the reconstructed field:
\begin{equation}
    SNR(k) = \frac{\langle |X_{true}|^2 \rangle}{\langle|\hat{X}-X_{true}|^2\rangle}
\end{equation}
Fig.~(\ref{fig:snrk_dm_de}) shows the SNR for the reconstructed dark matter (left) and electron densities (right) with both the marginalized (blue) and conditioned on the cosmological and astrophysical parameters (orange) models. We can see that the effect of global parameters on the mean SNR is more pronounced for the dark matter reconstruction. The conditional model also has smaller variance, as can be seen from the size of the corresponding $1\sigma$ regions.

\section{Model Input and Output Visualization}
\label{app:viz}

\begin{figure}[tbh!]
\centering
  \includegraphics[width=1.\linewidth]{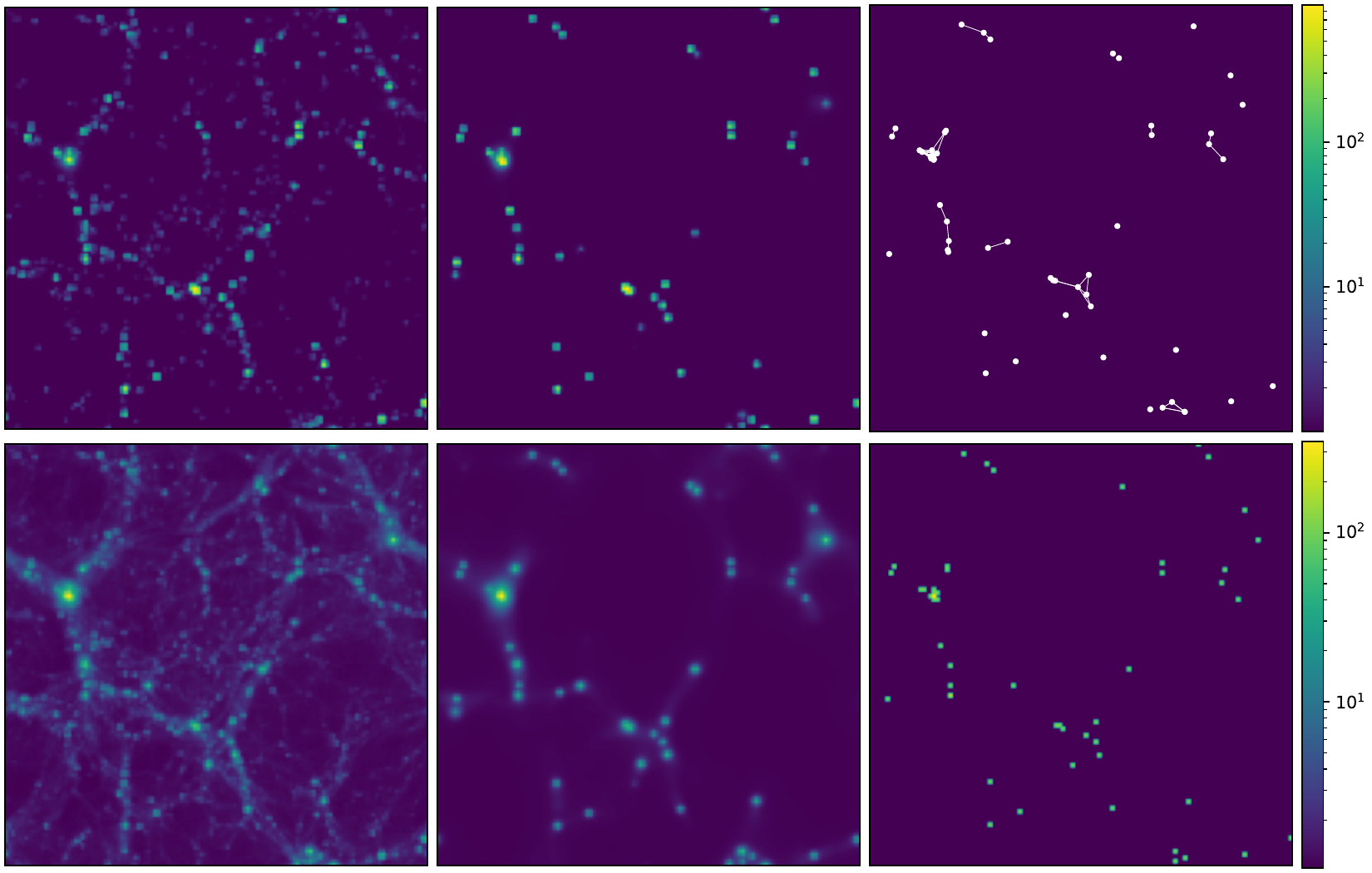}
  \caption{True (left) and predicted (middle) dark matter (bottom row) and electron (top row) densities in a $5\times25\times25\ (\texttt{Mpc/h})^3$ volume, averaged over the $x$ axis. The rightmost column shows the visualization of the input - galaxy cloud in the same region as a graph (top) and density field (bottom).}
  \label{fig:model_otp_compare2}
\end{figure}

\begin{figure}[tbh!]
\centering
  \includegraphics[width=1.\linewidth]{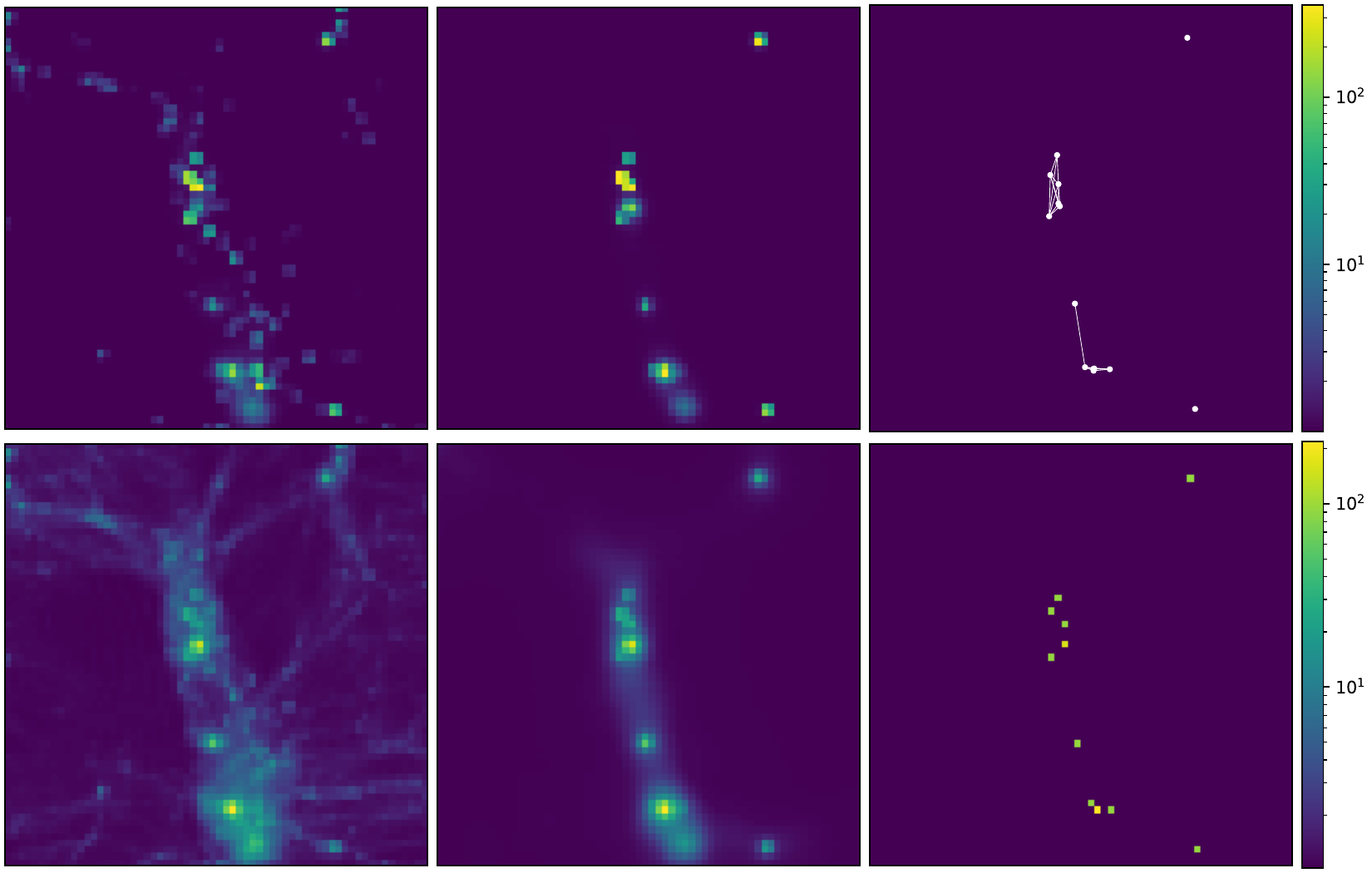}
  \caption{Same as Fig. \ref{fig:model_otp_compare2}, but in a zoomed-in volume of $5\times12.5\times12.5\ (\texttt{Mpc/h})^3$.}
  \label{fig:model_otp_compare1}
\end{figure}

We provide visual examples of the input and output data in Fig. \ref{fig:model_otp_compare2} and \ref{fig:model_otp_compare1}.

%%%%%%%%%%%%%%%%%%%%%%%%%%%%%%%%%%%%%%%%%%%%%%%%%%%%%%%%%%%%
\end{document}